# Fast Modeling L2 Cache Reuse Distance Histograms Using Combined Locality Information from Software Traces

Ming Ling, Jiancong Ge, Guangmin Wang
*National ASIC System Engineering Technology Research Center, Southeast University, Nanjing, China*
{trio, gejiancong, wangguangmin}@*seu.edu.cn*

*Abstract*—To mitigate the performance gap between CPU and the main memory, multi-level cache architectures are widely used in modern processors. Therefore, modeling the behaviors of the downstream caches becomes a critical part of the processor performance evaluation in the early stage of Design Space Exploration (DSE). In this paper, we propose a fast and accurate L2 cache reuse distance histogram model, which can be used to predict the behaviors of the multi-level cache architectures where the L1 cache uses the LRU replacement policy and the L2 cache uses LRU/Random replacement policies. We use the profiled L1 reuse distance histogram and two newly proposed metrics, namely the RST table and the Hit-RDH, that describing more detailed information of the software traces as the inputs. For a given L1 cache configuration, the profiling results can be reused for different configurations of the L2 cache. The output of our model is the L2 cache reuse distance histogram, based on which the L2 cache miss rates can be evaluated. We compare the L2 cache miss rates with the results from gem5 cycle-accurate simulations of 15 benchmarks chosen from SPEC CPU 2006 and 9 benchmarks from SPEC CPU 2017. The average absolute error is less than 5%, while the evaluation time for each L2 configuration can be sped up almost 30X for four L2 cache candidates.

## I. Introduction

As the speed gap between CPU and the main memory keeps increasing, multi-level caches are widely utilized in modern processors to improve the performance. Considering the high accuracies, in many early studies, researchers prefer to use cycle-accurate simulators to evaluate their designs [1]. However, as the complexity of architecture design spaces and the size of workloads are continuously growing, simulation time is becoming unacceptably high. Compared with cycle-accurate simulations, analytical models can provide faster performance estimations and architectural insights. They are normally based on software statistical information profiled from the workload traces. Since caches exploit memory accessing localities, Reuse Distance Histogram (RDH) and Stack Distance Histogram (SDH), which can be profiled by a binary instrumentation tool or other trace generators, have become the most important metrics for analytical models of caches with Random and LRU replacement policies [2][3]. However, because the profiled histograms merely reflect the memory accesses to the L1 caches, these two metrics cannot be directly used to predict behaviors in the downstream caches, i.e., the L2 and L3 caches.

In previous researches, there are many analytical models proposed to evaluate the performance of different cache architectures [4][5]. However, most of these models use pure probability formulas to predict the behaviors of a certain level cache, in which mathematical expectations and constant ratios are widely used instead of more accurate probability distributions. In fact, these values or probabilities used to describe the software trace information are closely related to the workloads. Using constant values in the model is inaccurate and cannot reflect the details of the distance histograms and thus, brings large errors. That is the reason why most previous models only provide the cache miss rates of the target caches without more insightful distance histograms [2][6][7]. Furthermore, their models usually require the replacement policy of the target cache being LRU. Because as long as the proportion of the references with stack distances less than the associativity is accurate, the miss rate of an LRU cache can be calculated easily and accurately. Therefore, the accuracies of these miss rate models are not sensitive to the precisions of the distance histograms. For example, Fig. 1 gives the L1 SDH of *cactusADM* in SPEC CPU 2006 predicted by StatStack [5] from the L1 RDH. Although StatStack can give the accurate prediction of the miss rate, the distance histogram shown in Fig. 1 is not accurate enough. The predicted histogram has a high peak around stack distance 25, but the peak appears around 50 in the actual distance histogram with a significantly lower height. However, because the proportion of references with the stack distance less than the associativity (4-way, in this example) is similar, the miss rates calculated by two distance histograms are also similar (the actual miss rate is 2.47% while the predicted miss rate is 2.02%).

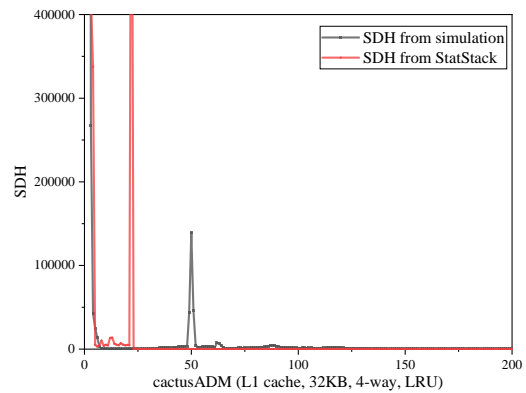

Fig. 1. L1 SDH from StatStack and L1 SDH from simulations

In addition, the profiled information used in previous models is either RDH or SDH, without any combinations of these two metrics. However, a single RDH or SDH is incapable to describe all information of the software traces. RDH cannot be directly used to analyze the "filter effect" of the L1 LRU cache because only the stack distance of a reference can be utilized to determine whether it is missing in the LRU cache. SDH, on the other hand, filters identical references in a reuse epoch and cannot be directly used to derive the L2 RDH.

Based on the above considerations, it is not enough to analyze the L2 RDH merely through the L1 RDH or the L1 SDH due to their incompleteness. In this paper, we propose two new metrics, namely the Reuse-and-Stack-Transfer (RST) table and Hit-RDH, to describe more detailed information of the software trace from CPU. Our model, which takes L1 RDH, RST table, Hit-RDH and corresponding cache parameters as inputs, can infer the accurate L2 RDH without detailed simulations of the cache architecture. With the help of StatCache [4] and StatStack [5], the L2 RDH can be used to calculate the L2 cache miss rates for L2 Random cache and L2 LRU cache, respectively. For a given L1 cache configuration, the profiled RST table and Hit-RDH can be reused for different L2 cache configurations. The profiling process only needs to be re-run when the L1 configuration changed.

Our work improves the related works in the following two aspects:

- Proposing two new metrics to describe more detailed information of software traces, which can be profiled from conventional method without significant time and space overhead.
- Providing an analytical model, which is more accurate and more time efficient than prior probability models, to derive the L2 RDH from L1 RDH that can be used to predict the corresponding L2 cache miss rate.

The rest of the paper is organized as follows: Section 2 introduces the related works. Section 3 reviews the background of the analytical model for caches. Section 4 introduces two new metrics for describing the information of software traces. In Section 5, we propose an analytical model equation set to predict the L2 RDH. In Section 6, we give the solution for the scenarios when the L1 cache and the L2 cache have different number of cache sets. The evaluation results of our model are exhibited in Section 7. Section 8 concludes this paper.

## II. RELATED WORKS

Prior researches have provided some very useful metrics for us to model the behaviors of caches. Kristof et. al [8] proposed reuse distances to describe the behaviors of cache. They also provided some directions about new optimizations focused on cache performance. Yutan Zhong et. al [7] proposed an approximate reuse distance measurement and statistically predicted the locality of programs.

Regarding cache miss rate estimations, the related cache modeling methodologies can be categorized into two parts. The first part is for the models that focus on a certain cache level, especially the L1 cache. Erick Berg et al. [4] presented an analytical model, StatCache, to estimate the L1 cache misses with the Random replacement policy. This model is fed with the RDH profiled from the memory references. It is basically composed of two equations. By solving the equation set, we can get the miss rate of the Random cache. David Eklov et al. [5] developed StatStack that converts the RDH into the SDH from which the L1 LRU cache misses can be predicted. Xiaoyue Pan et al. [9] provided a framework based on the Markov chain to predict the cache misses under three replacement policies, namely Random, LRU and PLRU. For the out-of-order processors, K Ji et al. [10] used artificial neural networks to address the effects of the stack distance migration that caused by out-of-order executions.

The second part is for the models for multi-level cache architectures. K Ji et al. [11] [12] used the L1 cache SDH to predict the multi-level cache misses based on a total probability formula. It considers all the possible situations of distance distributions in a reuse epoch, which makes the complexity of the algorithm extremely high. Jasmine Madonna S et al. [13] proposed an analytical model to calculate the L2 cache miss rate based on the analysis of the influence of inclusive/exclusive relationship between the L1 cache and the L2 cache. Nevertheless, the output of their model only gives the L2 miss rate without the more insightful L2 RDH. In addition, these models are only valid for the LRU-LRU multi-level cache architectures. Venkatesh T G et al. [14] and David Eklov et al. [15] have proposed methods to construct the L2 RDH of the shared cache. However, the inputs of their models are profiled from the direct upstream caches instead of the CPU, which can only be obtained via the time-consuming simulations of the multi-level cache architecture.

In our previous work [16], we put forward a new metric to describe the characteristics of software traces, by which the proposed model can get a relatively accurate L2 RDH under the low L1 miss rates scenarios. However, the model requires that the L1 cache and the L2 cache have the same number of cache sets, which limits the usage scope of the model in real processor architectures.

## III. BACKGROUND

Before we introduce the new metrics and our model, we first need to review some basic terminologies and backgrounds used in our following discussions.

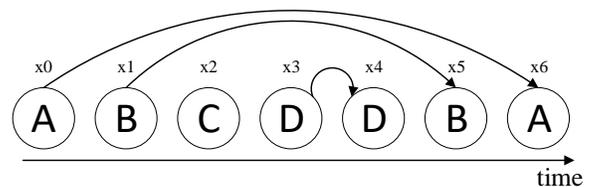

Fig. 2.  A Reuse Epoch

*Reuse distance*: The reuse distance is the number of references between two consecutive references accessing to the same cache line [5]. If the cache line is accessed for the first time, its reuse distance is defined as infinite. For example, the reuse distance of the second A in Fig. 2 is 5 (there are five references, $x_1$, $x_2$, $x_3$, $x_4$ and $x_5$, between the two 'A's).

*Stack distance*[1]: The stack distance is the number of distinct references between two consecutive references accessing to the same cache line [5]. In Fig. 2, the stack distance of the second A is 3 (There are three distinct references, $x_2$, $x_4$ and $x_5$, between the two 'A's).

*Reuse/Stack Distance Histogram (RDH/SDH)*: The RDH/SDH records the numbers of references for each reuse/stack distance in the memory traces.

*Reuse epoch*: A reuse epoch is the cache accessing history between two consecutive references accessing the same cache line, such as the reuse epochs formed by the two 'A's ($x_1$, $x_2$, $x_3$, $x_4$ and $x_5$) and the two 'B's ($x_2$, $x_3$ and $x_4$) in Fig. 2.

*StatCache*: StatCache is proposed by Erik Berg and Erik Hagersten [4], which is mainly composed of two equations, shown in Eq. (1). By solving the equation set, it can predict the miss rate of a Random cache.

$$\begin{cases} f(n) = 1 - \left(1 - \dfrac{1}{assoc}\right)^n \\ RN \approx h(1)f(R) + h(2)f(2R) + h(3)f(3R) + \cdots \end{cases} \quad (1)$$

In this equation, $f(n)$ means the possibility that one reference has already been evicted from the Random cache after $n$ misses are generated. Because the cache line is replaced randomly, the possibility that one certain cache line is evicted is $\dfrac{1}{assoc}$. After $n$ cache misses, the possibility that the reference is still in the cache set is $\left(1 - \dfrac{1}{assoc}\right)^n$. In the second equation, $R$ is the miss rate of the Random cache while $N$ is the total number of memory references. The term $h(x)$ represents the number of references with the reuse distance of $x$. The left side of the second equation is the total number of misses in the Random cache. On the right side of the second equation, every factor $h(x)f(xR)$ means the number of misses generated by the references with reuse distance $x$. After solving the equation set, the miss rate of the Random cache can be roughly estimated.

*StatStack*: To predict the miss rate of an LRU cache with RDH, David Eklov and Erik Hagersten proposed StatStack [5]. StatStack derives an expected stack distance histogram, $ES(r)$, from the reuse distance histogram. Although the expected stack distance histogram is not the actual SDH, it is accurate enough to predict the miss rate of the LRU cache. Eq. (2) shows the way to calculate the expected stack distance histogram.

$$ES(r) = \dfrac{1}{n_r} \sum_{x_i \in T(r)} \sum_{j=1}^{r} 1\left(\vec{R}\left(x_{i-j}\right) > j\right) \quad (2)$$

In Eq. (2), $\left(\vec{R}\left(x_{i-j}\right) > j\right)$ means the ratio that in the reuse epoch of reference $i$, the forward reuse distance of the $j^{th}$ reference larger than $j$ which should be counted as the stack history. After accumulating all the values of $1\left(\vec{R}\left(x_{i-j}\right) > j\right)$ in the reuse epoch, the expected stack distance can be calculated. To predict the miss rate, we just use the number of references whose expected stack distance is larger than the associativity dividing the total number of references.

*Extending StatCache and StatStack to the set-associativity caches*: In this paper, StatCache and StatStack are used to calculate L2 cache miss rates from the predicted L2 RDHs. Considering both tools are designed for full-associativity caches, we need to extend them into set-associativity scenarios. In set-associativity caches, when we calculate the reuse distance/stack distance, only the references indexed to the same cache set will be counted because the references indexed to different cache set will not occupy the cache line in other cache set. For example, as shown in Fig. 3, reference A, C, D are indexed to set 0 and reference B is indexed to set 1. We notice that no matter how many references of B appeared in the reuse epoch of reference A, they will never occupy any cache line in set 0 and, thus, the miss number of A, C, D will not be affected. After collecting the RDH/SDH for each cache set (sometimes called set RDH, SRDH or set SDH, SSDH), we accumulate the RDHs/SDHs of all cache sets linearly and get the overall RDH/SDH [10], based on which StatCache and StatStack could be utilized to calculate the miss rate. Obviously, the need to extract reuse distances and stack distances for each cache set by individual linked-lists is more complex than extracting such information from a fully-associative cache, in which only one or two global linked-lists are required. However, considering the number of cache sets in an L1 cache is relatively small and different memory references are distributed into each individual linked-list, the space and time complexities of our approach can be limited in a reasonable bound.

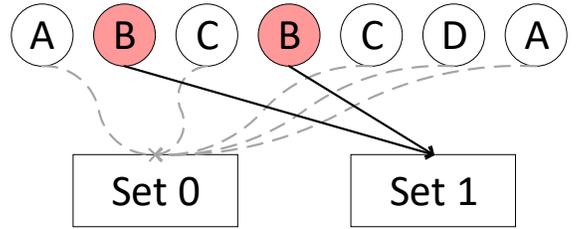

Fig. 3.  References indexed to different cache sets do not affect each other when extracting the reuse distances or stack distances

*Overview of the working flow*: Fig. 4 shows the estimation framework of our model. We use gem5 [17] working under *AtomicSimpleCPU* mode as our trace profiler[2]. First, the applications are running on gem5 and our inserted profiling code will collect the RDHs/SDHs, RST tables and Hit-RDHs of the applications. Because the profiling results can be reused by different L2 cache configurations, it is not necessary to re-conduct this procedure for a given L1 configuration. Second, our model takes the profiled trace information and L1/L2 cache configurations as the inputs to predict the L2 RDH. Based on the replacement policy the L2 cache applied, we leverage StatCache or StatStack to calculate the miss rates for L2 Random or L2 LRU caches, respectively.

---

[1]Some prior researchers also define the stack distance as "reuse distance" and the reuse distance as "time distance". However, we apply the names and definitions used in [4] and [5].

[2]Theoretically, the software trace can be generated by any trace generator, like Pin. However, we only extract memory locality information and do not need to store the huge software traces. Therefore, by inserting the profiling code into gem5, we can obtain the locality information during application execution.

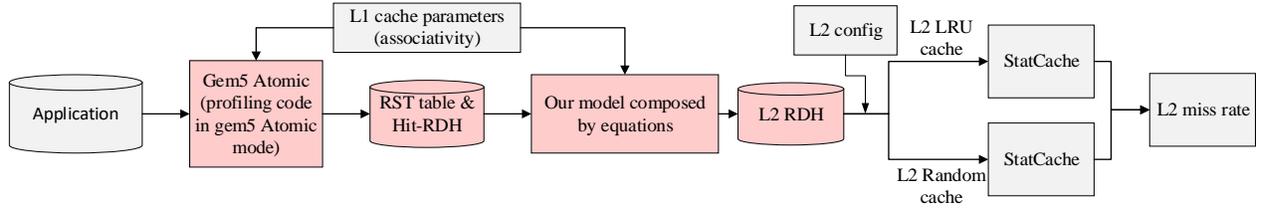

Fig. 4. Overview of the Estimation Framework

## IV. RST TABLE AND HIT-RDH

In previous studies, the stack distance theory [5] is the most important method to analyze the performance of the LRU cache. For the Random caches, StatCache [4] is used to calculate the miss rate based on the RDH. These methods only output the cache miss rates and none of them analyze the "filter effect" of the L1 cache. The information of the software traces will be changed after accessing the L1 cache because some memory references may hit in the L1 cache and be filtered from accessing the L2 cache. However, the memory traces, or the distance histograms, which are normally profiled by a binary instrumentation tool, only reflect the memory accesses to, instead of from, the L1 cache. Therefore, prior models are very limited in the L2 cache modeling.

To describe the "filter effect" of the L1 cache, we propose two new metrics, namely the Reuse-and-Stack-Transfer (RST) table and Hit-RDH in this paper. RST table is a two-dimensional matrix, which records information of the RDH and the SDH in a given trace profiling interval in the L1 cache (in this paper, we extract the statistical information, e.g., the SDH, RDH, RST table and Hit-RDH, for every 500 million instructions). As the example shown in Fig. 5, every element in the RST table contains the relationship between the reuse distance and the stack distance. The red circle $RST[4][1]$ in Fig. 5 represents there are 320 references in this interval with the reuse distance of 4 and the stack distance of 1. Moreover, for given references with the reuse distance of $i$, we use Eq. (3) to calculate the normalized RST table, called $Prs$. We define each element $Prs[i][j]$ as the probability that the references have the stack distance of $j$, which is the proportion of $RST[i][j]$ in the whole $i^{th}$ row. For instance, as shown in Fig. 5, the red circle in the normalized RST table means that in all references with the reuse distance of 4, 76% references have the stack distance of 1.

$$Prs[i][j] = \frac{RST[i][j]}{\sum_{k=0}^{i} RST[i][k]} \quad (3)$$

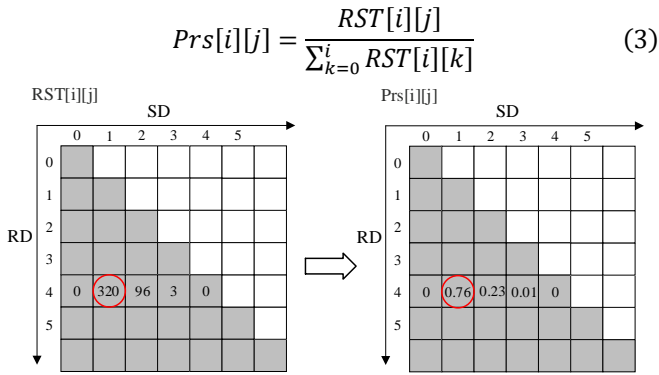

Fig. 5. The RST table and Normalized RST table ($Prs$)

Another metric introduced in this paper, called Hit-RDH, is also a two-dimensional matrix. Fig. 6 shows an example of Hit-RDH. The red circle in Fig. 6 means that in all the reuse epochs with the reuse distance of 4, the number of reuse epochs that have 2 references hitting in the L1 cache is 310. In other words, there are 310 reuse epochs whose reuse distance are 4 and in each of them there are 2 references hit in the L1 cache. By Eq. (4), we can also get the normalized Hit-RDH, called $P_{N_{hit}}$, as shown in Fig. 6. $P_{N_{hit}}[i][j]$ is the proportion of the $Hit_{RDH}[i][j]$ in the whole $i^{th}$ row.

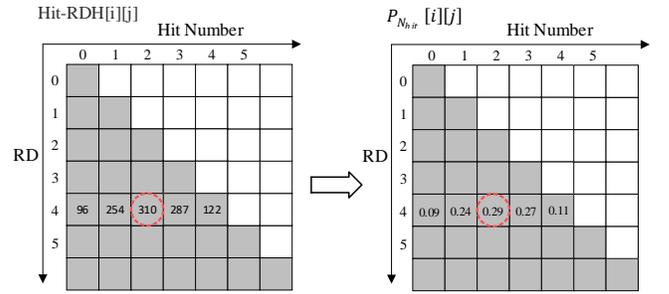

Fig. 6. Hit-RDH and Normalized Hit-RDH ($P_{N_{hit}}$)

$$P_{N_{hit}}[rd][n] = \frac{Hit_{RDH}[rd][n]}{\sum_{k=0}^{rd} Hit_{RDH}[rd][k]} \quad (4)$$

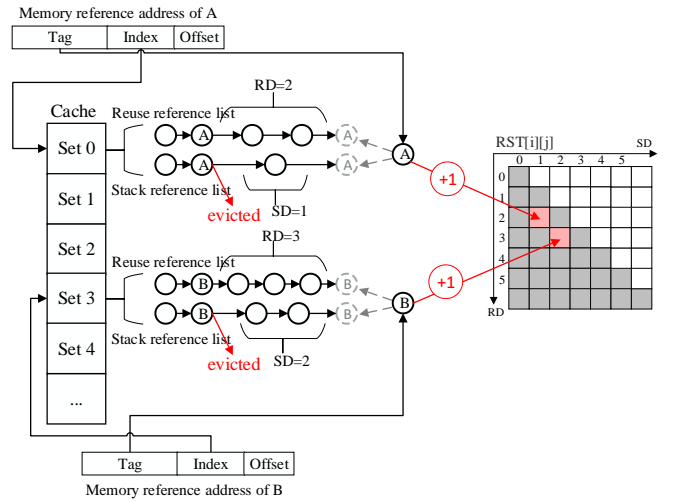

Fig. 7. Reference lists used to calculate the reuse distance and the stack distance

To construct RST and Hit-RDH tables for a set-associative cache, we need to maintain two linked-lists to record the reuse/stack history of each memory access that indexed to every individual cache set. Every memory reference contains the accessing address of the target memory. We first extract the *index* field, which can be determined by the number of cache sets and the cache line size. In our experiments, for example, if the L1 cache is equipped with 32KB and 4 way-associative with 64 bytes per cache line, the *offset* field occupies 6 bits and the *index* field takes 7 bits (128 sets with 4 cache lines per set and 64 bytes per cache line). The *tag* can be extracted easily by masking the *index* bits and *offset* bits.

When a memory reference *A* comes, as Fig. 7 shows, the index bits are firstly used to address the corresponding set linked-lists, while the extracted *tag* is pushed to ends of the reuse reference list and the stack reference list. When we calculate the reuse distance, we just count the number of references between these two references of *A* in the reuse reference list. If we want to get the stack distance, we count the distance between two '*A*'s in the stack reference list. Then, we delete the reference *A* found in the recent history in stack reference list (evicting the previous reference *A* we found in the stack reference list), because there is no duplicate reference in the stack history. By using this method, we can get the reuse distance and the stack distance for each memory reference. To construct the RST and Hit-RDH tables, we just need to increase the value by one in the corresponding element of each table for each coming reference, as shown in Fig. 7. For example, if a memory reference indexed to set0 has a set reuse distance $i$ and set stack distance $j$, we increase the value of $RST[i][j]$ by one; while another memory reference indexed to set1 has a set reuse distance $l$ and set stack distance $m$, we update the element of $RST[l][m]$ with a new value by adding one. This RST and Hit-RDH tables updating procedure is just an attached processing when profiling the RDH and SDH. Thus, the extra time overhead of maintaining these two tables is negligible. More details about the extraction of the reuse distance and the stack distance could be found in [11][18][3]. It is also worth to note that real memory accessing stream arriving at the L1 cache may be different from the accessing order seen by a binary instruction tool or the gem5 AtomicSimpleCPU simulation. The memory accessing order could be changed by the aggressive Out-of-Order execution mechanisms, such as dynamic instruction scheduling, memory level parallelism (MLP) offered by the MSHRs, prefetching, speculative loads and load in stores. In such cases, the extracted stack distance and/or the reuse distance for a given memory access could be changed by the aforementioned hardware components. To precisely quantify these influences on the SDH or RDH that extracted by software trace profiling without detailed simulations is extremely difficult, if not impossible. Our prior work has tried to build an empirical model based on artificial neural networks (ANN) to quantitatively analyze the impacts of these factors [10][18]. Nevertheless, the detailed discussion about the effects of Out-of-order mechanisms on the SDH and RDH is out of this paper's scope. Therefore, we will ignore these effects in our paper.

Both RST table and Hit-RDH table are two-dimensional arrays with every element a 4-byte integer. To make a tradeoff between space/time overheads and accuracies, we choose to cut off the profiled L1 reuse/stack distances at 1024, which is also applied in the work of [10][11][12][18]. In [19], we find that most reuse distances (>95%) of all benchmarks in SPEC CPU2006 are less than 32KB. Because we use the cache line aligned address to extract reuse/stack distances, the cutting off distance at 1024 covers a memory range of 64KB (1024*64 Bytes, given the cache line size in this paper is 64 Bytes), which is large enough for our study. In addition, by accumulating the appearance numbers of memory references with reuse/stack distances larger than 1024 to the terms of RDH (1024) and SDH (1024), we keep the total memory reference number same as that of software traces. Therefore, the space cost of RST table and Hit-RDH tables is 8MB (2 * 1024 * 1024 * 4 bytes), which is relatively small considering the physical memory capacity of the profiling platform is 8GB.

## V. L2 CACHE RDH MODEL

To simplify our discussion of the L1 cache "filter effect", we first assume that: (1) the L1 cache equips with the LRU replacement policy; (2) the L1 cache and the L2 cache have the same number of sets (we will extend our model to different L1/L2 sets in Section VI); (3) although the L2 cache is shared by the L1 D-Cache and L1 I-Cache, like in [11][13], we ignore the influences from L1 I-Cache considering its small impacts on the L2 cache behavior. Based on the stack distance theory, if the stack distance of a reference is larger than or equal to the LRU-cache associativity, the reference must be a cache miss. In other words, the references with stack distance smaller than the L1 associativity will hit in the L1 cache without accessing the L2 cache. However, we cannot derive the L2 RDH directly from the L1 SDH because it does not record the information about how many times a certain reference is reused in a reuse epoch. For example, Fig. 8 shows a reuse epoch of $A$. We assume that the L1 associativity is 2. References $x_0$ and $x_6$ construct the reuse epoch of $A$. The references $x_3$ will hit in the L1 cache because the stack distance of $x_3$ is 1, which is less than the L1 associativity. Although reference $C$ has appeared twice (reference $x_2$ and reference $x_5$), reference $x_5$ will still be a miss because its stack distance is 2. Based on the above analysis, only reference $x_3$ hits in the L1 cache and others will be leaked to the L2 cache. If the references $x_0$ and $x_6$ miss in the L1 cache, the reuse distance of the second reference $A(x_6)$ in the L2 cache is counted by the references in the reuse epoch missing in the L1 cache, i.e., the reuse distance of the second A in the L2 cache is 4.

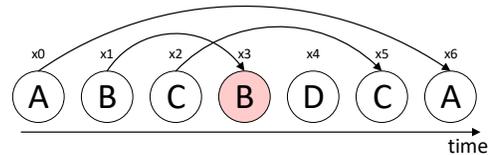

Fig. 8. An example for the reuse distance and the stack distance in a 2-way L1 cache

---

[3] The main target of this paper is the study on how to derive the L2 RDH from L1 RDH/SDH without time-consuming full simulations. Therefore, we do not implement any sampling technique nor advanced RDH/SDH extracting methods, e.g., tree-based algorithm, in this paper. However, these techniques are orthogonal to the topic of this paper and can be integrated into our work easily, such as Reuse Distance Vector (RDV) introduced in [20].

If we want to predict the L2 RDH from the distance information of the L1 cache, for a reuse epoch of a reference in the L1 cache, we have to know how many references hit (or miss) in the L1 cache. Then, we can calculate the number of references leaked into the L2 cache and the corresponding reuse distance of the L2 cache. However, the L1 stack distance of a reference is not the total number of the references in its reuse epoch, which means we are unable to calculate the reuse distance of the L2 cache by judging how many references are missing in the L1 cache. For example, the stack distance of the second $A$ in Fig. 8 is 3 (there are 3 distinct references, i.e., B, C and D, between two $A$s), and there is only one reference, $x_3$, hitting in the L1 cache. However, we cannot get the L2 reuse distance of the second A, which is 4 in this case, merely based on the stack distance information. Therefore, the only way to calculate the reuse distance of L2 cache is that we get the reuse distance of the reuse epoch of a reference and determine how many references in this reuse epoch are misses.

In our model, for a reference in the L1 cache, we first use the $Prs$ table to get a distribution of its possible stack distances. By the definition of $Prs$ table, we know $RDH(i) \times Prs[i][j]$ denotes how many references with the reuse distance of $i$ have the stack distance of $j$, where $RDH(i)$ is the appearance number of references with the L1 reuse distance of $i$ and $Prs[i][j]$ can be considered as the ratio that the references with the reuse distance of $i$ and the stack distance of $j$. If $j$ is smaller than the L1 associativity, $RDH(i) \times Prs[i][j]$ references will hit in the L1 cache and not be leaked into the L2 cache. We subtract the number of the L1 hit references from each $RDH(i)$, as Eq. (5) shows, to get an intermediate histogram $MissRDH$.

$$MissRDH(i) = RDH(i) \times \left(1 - \sum_{j=0}^{L1\ Assoc-1} Prs[i][j]\right) \quad (5)$$

Note that $MissRDH$ is not the L2 RDH yet, which only reduces the number of the references of each reuse distance in the L1 RDH, as Step 1 in Fig. 9 shows. For each reuse epoch, some references may hit in the L1 cache. Thus, when the reuse epoch leaked into the L2 cache, its reuse distance might be decreased and it should be counted to a lower reuse distance bar like the blue boxes in Step 2 in Fig. 9. Therefore, to get the L2 RDH, we still need to adjust the number of each reuse distance.

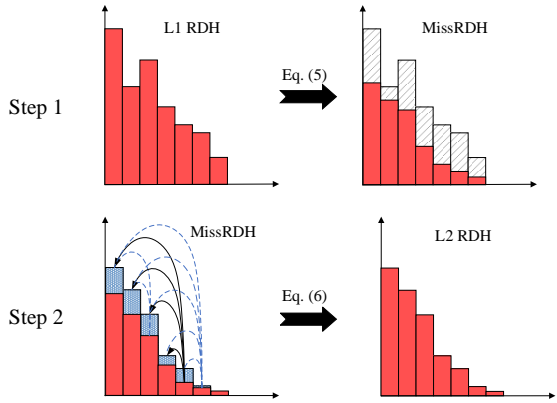

Fig. 9. Steps from the L1 RDH to the L2 RDH

After Step 1 in Fig. 9, the value of each bar in the $MissRDH$ is the number of references with the original corresponding reuse distances that are misses in the L1 cache, but the actual reuse distances of these references in the L2 cache have not been adjusted. In Step 2, we use the proportion $P_{N_{hit}}$ to adjust $MissRDH$. By the definition, $P_{N_{hit}}[rd][n]$ means that in all the reuse epochs with the reuse distance of $rd$, the proportion of how many reuse epochs have $n$ hit references in their reuse epochs. If the reuse distance of a reuse epoch in the L1 cache is $rd$ and while the reuse distance in the L2 cache is $i$, that means $rd - i$ references in the reuse epoch hit in the L1 cache and the ratio of these references is $P_{N_{hit}}[rd][rd-i]$. Eq. (6) shows the way to get the L2 RDH from $MissRDH$. In this equation, $MissRDH(rd) \times P_{N_{hit}}[rd][rd-i]$ means how many memory references with L1 reuse distance $rd$ have been moved, or migrated, to a given L2 reuse distance bar $i$ because there are averagely $rd - i$ references are L1 hits in each of the reuse epoch. By accumulating all the migrated references from each higher bar $(rd > i)$ in $MissRDH(rd)$, we can obtain the adjusted $L2RDH(i)$. Repeating this process from $i = 0$ to infinite (because we set the cutting off reuse distance at 1024, the process actually will be done for 1024 times), we can construct the whole $L2RDH(i)$, where $i$ is from 0 to 1024. For example, the number of references miss in the L1 cache construct the reuse distance of the reuse epoch in the L2 cache, as Fig.10 shows. In Fig. 10, we assume that the references at both ends (two references of $A$) are missing in the L1 cache. Reference $x_3$ and reference $x_6$ hit in the L1 cache and will not access the L2 cache. Other references $(x_1, x_2, x_4, x_5)$ are missing in the L1 cache and accessing the L2 cache, which construct the reuse epoch in the L2 cache.

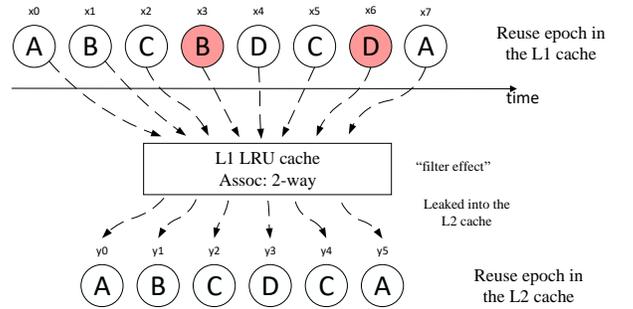

Fig. 10. The references miss in the L1 cache construct the reuse epoch in the L2 cache

$$L2RDH(i) = \sum_{rd=i+1}^{\infty} MissRDH(rd) \times P_{N_{hit}}[rd][rd-i] \quad (6)$$

## VI. L1 CACHE AND L2 CACHE HAVE DIFFERENT NUMBER OF SETS

In real hardware, the number of sets in the L1 cache and in the L2 cache are usually different. We assume that the size of both L1 and L2 cache line is 64 bytes. If the capacity of the L1 cache is 32KB and the associativity is 2, the number of sets in the L1 cache is 256. If the capacity of the L2 cache is 512KB and the associativity is 8, the number of sets in the L2 cache is 1024. In this case, the sets number of the L2 cache is four times of the L1 cache, which means two different references indexed to the same set in the L1 cache may be indexed to two different sets in the L2 cache. We define $P_{same}$ as the probability that two different references are indexed to the same cache set in the L1 cache and in the L2 cache. $P_{same}$ can be calculated by Eq. (7) [11] [13].

$$P_{same} = \frac{S_{L1}}{S_{L2}} \quad (7)$$

In Eq. (7), $S_{L1}$ is the number of sets in the L1 cache and $S_{L2}$ is the number of sets in the L2 cache. After Step 1 and Step 2 in Fig. 9, we still need to adjust the L2RDH, because the references in a reuse epoch may be indexed to different sets in the L2 cache.

In the bottom of the Fig. 10, it is the reuse epoch predicted by the above method without considering the situation that the references are indexed to different sets in the L2 cache. If the L1 cache and the L2 cache have different number of sets, some references in this reuse epoch may be indexed to other cache sets with the reference $A$, which should not be counted as the reuse distance for the reference $A$ in the L2 cache. To simplify the discussion, the RDH predicted without considering the different set number of two-level caches in Fig. 9 is named as L2RDH, while the real RDH of the L2 cache is represented as RealL2RDH.

Fig. 11 gives an example of all cases when there are two references being indexed to the same set with references $A$ in the L2 cache with the reuse distance of 4 in L2RDH. If one reference in this reuse epoch is indexed to a different set with reference $A$ in the L2 cache, its probability is $1 - P_{same}$. We assume that the memory accessing obeys the uniform distribution[4] and the $ReadL2RDH(2)$ contributed by this reuse epoch is $1 \times C_4^2 \times (1 - P_{same})^2 \times (P_{same})^2$, in which symbol $C$ means combination and $C_4^2$ means the number of combinations of choosing two items from a set with 4 elements.

Similarly, we can extend the example into a general case, in which the reuse distance of the reuse epoch in L2RDH is $rd1$ and the reuse distance in RealL2RDH is $rd2$. The $RealL2RDH(rd2)$ can be calculated by Eq. (8).

$$RealL2RDH(rd2) = \sum_{rd1=rd2}^{\infty} L2RDH(rd1) \times C_{rd1}^{rd2} \times (P_{same})^{rd2} \times (1 - P_{same})^{rd1-rd2} \quad (8)$$

We can notice that in Eq. (8), if the L1 cache and L2 cache have the same number of cache sets, $P_{same}$ will be 1. Therefore, only when $rd1$ equals to $rd2$, $(P_{same})^{rd2}$ and $(1 - P_{same})^{rd1-rd2}$ become 1 and $RealL2RDH(rd2)$ equals to $L2RDH(rd1)$. In other cases, $(1 - P_{same})^{rd1-rd2}$ is zero and the reuse distance of the references will not be adjusted. So, Eq. (6) is a special case of Eq. (8).

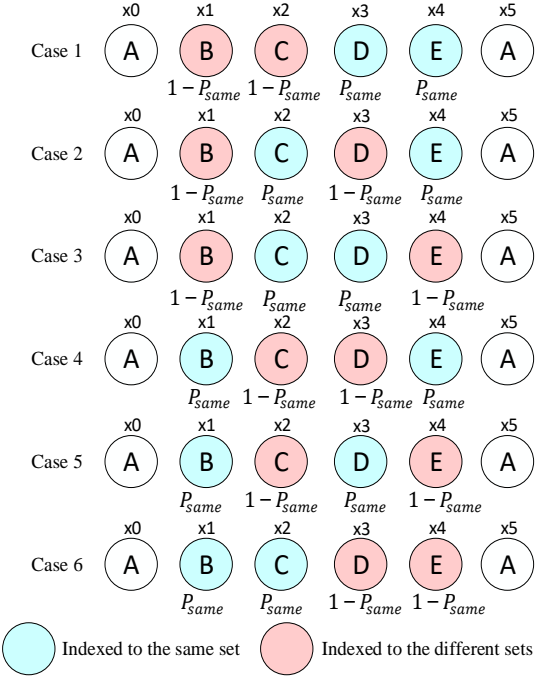

Fig. 11. All cases when the reuse distance of reference $A$ is 2 in the L2 cache

VII. EVALUATION

In this section, we use gem5 simulator [17] to evaluate our model. The cache architecture used in the experiments has two levels and the detailed configurations are shown in Table 1. Fifteen benchmarks chosen from SPEC CPU 2006 benchmarks are used to evaluate our proposal with the *ref* input set[5] (shown in Table. 2). We also evaluate our model with nine applications from the latest SPEC CPU 2017 benchmarks. We compared the L2 RDH curves and the L2 miss rates with the outputs of the cycle-accurate gem5 simulations. The L2 cache miss rates of our model can be evaluated by StatCache and StatStack depending on the replacement policy applied.

Table. 1 Multi-level Cache Hardware Configuration

| ISA | | X86 | |
|---|---|---|---|
| Pipeline | | 7-Stages, Out-of-Order | |
| L1 16KB 2-way LRU | | L1 32KB 4-way LRU | |
| **Config1** | L2 64KB 8-way Random | **Config2** | L2 128KB 16-way Random |
| **Config3** | L2 64KB 8-way LRU | **Config4** | L2 128KB 16-way LRU |
| **Config5** | L2 128KB 8-way Random | **Config6** | L2 512KB 16-way Random |
| **Config7** | L2 128KB 8-way LRU | **Config8** | L2 512KB 16-way LRU |

In this paper, we use average absolute error to estimate the accuracy of our model. The error for each benchmark is calculated by $|missrate_{model} - missrate_{gem5}|$, in which $missrate_{model}$ and $missrate_{gem5}$ mean the miss rates from our model and gem5, respectively. The average error in a fixed configuration can be estimated by Eq. (9). Eq. (9) is for comparing the errors between the miss rate calculated by our model and simulated by gem5 for each benchmark in one fixed configuration of multi-level caches, in which $N_{bench}$ means the number of benchmarks we used. Eq. (10) calculates the total average error by dividing the sum of all errors of $error_{\$config}$ with the number of configurations in our experiments, in which $N_{\$config}$ means the number of different L2 cache configurations. The cache configurations we used in our experiments are also shown in Table. 1.

---

[4] Although real memory accesses may follow a non-uniformed distribution, to quantify this distribution could be very complex and time-consuming. Similar to the works in [11] and [13], we accept the assumption of a uniform distribution to simplify our discussion. The following evaluation results also validate this assumption.

[5] Except for *mcf* and *gromacs*, which using the *test* input set.

$$error_{\$config} = \frac{\sum_{N_{bench}} |missrate_{model} - missrate_{gem5}|}{N_{bench}} \quad (9)$$

$$error_{total} = \frac{\sum_{N_{\$config}} error_{\$config}}{N_{\$config}} \quad (10)$$

We also proposed a metric, Histogram Error (*HE*), to describe the difference between two RDHs. Eq. (11) gives the way to calculate *HE*. In this equation, $RDH_m(i)$ means the result from our model and $RDH_{gt}(i)$ is the ground truth RDH from gem5 simulations.

$$HE = \frac{\sum |RDH_m(i) - RDH_{gt}(i)|}{\sum RDH_{gt}(i)} \quad (11)$$

Fig. 12 and Fig. 13 show the comparisons of the L2 RDHs of four benchmarks from gem5 simulations and our model under two different cache configurations (Config1 and Config2). The x-axis is for the reuse distance and the y-axis is for the number of references of the corresponding reuse distance in the L2 cache. We also label the corresponding *HE* of each sub-figure calculated from Eq. (11) in the figures. We simulate 500 million instructions for each benchmark. In these two configurations, the L1 cache and the L2 cache have the same number of sets. From these figures, we can find that the outputs of our model reflect most of the characteristics of the L2 RDH curves from gem5 simulations. The average *HE* in Fig. 12 and Fig. 13 is around $10^{-4}$, which means the average accumulated error of each RDH bin takes 0.07% of all memory accesses. Because we set the cutting off reuse distance as 1024 and accumulate references with larger distances into the bin of 1024, we can find both the simulation and the model curves have high peaks at the distance of 1024, which are mainly caused by first references to new addresses. Moreover, because we only simulate 500 million instructions, the proportions of these first references are relatively high. Fig. 14 gives the comparison results of L2 cache miss rates of all fifteen benchmarks from SPEC CPU 2006 and nine benchmarks from SPEC CPU 2017 under four configurations (Config1, Config2, Config3 and Config4). Random replacement is applied in the L2 caches for config1 and config2. We use StatCache to evaluate the miss rates of the L2 Random caches. LRU replacement policy is used in the L2 cache config3 and config4. StatStack is used to calculate the miss rates of the L2 LRU caches. In order to rule out the errors caused by StatCache and StatStack, we compare the calculated miss rates by our model with the miss rates calculated from the L2 RDH extracted from gem5 simulations. The average absolute errors of benchmarks from SPEC CPU 2006 under four scenarios are 1.94%, 3.82%, 1.82% and 3.86%, respectively. Meanwhile, the average absolute errors of benchmarks from SPEC CPU 2017 with cache Config3 and Config4 are 1.62% and 5.71%.

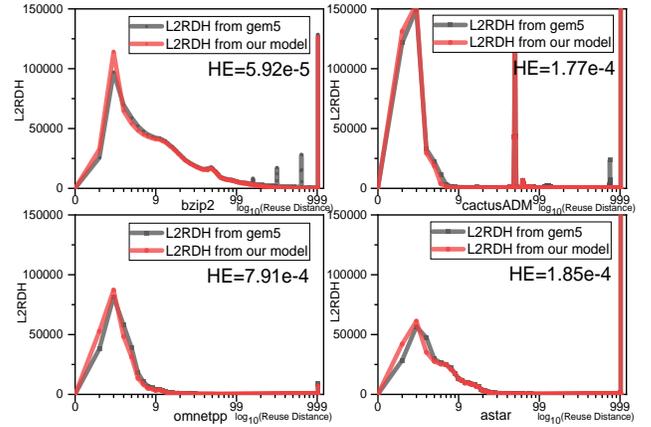

Fig. 12. Comparisons of the L2 RDH between gem5 simulation results and our model (L1 16KB 2-way, L2 64KB 8-way)

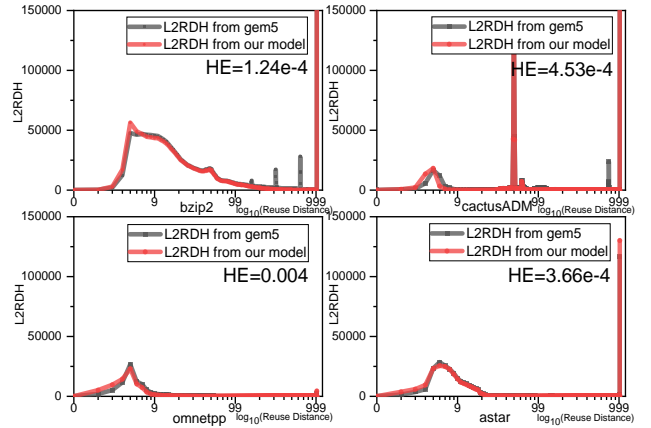

Fig. 13. Comparisons of the L2 RDH between gem5 simulation results and our model (L1 32KB 4-way, L2 128KB 16-way)

From Fig. 14, we can find a strange phenomenon that the L2 cache miss rates under the config2 and config4 are higher than those under the config1 and config3 (403, 429, 464), in which the cache size and associativity in config2 and config4 are larger than those of config1 and config3. This is because the traces poured into the L2 caches are different caused by the L1 cache "filtered effect" (L1 cache configurations are different). In config2 and config4, the L1 caches have larger size and associativity, which could utilize the memory reference locality better. The references with good locality are more easily hitting in the L1 cache with larger size and associativity. On the contrary, the locality of the references missing in the L1 cache and leaked into the L2 cache are not so good (many of these references are cold misses). That is the reason why the miss rates of the L2 cache under config2 and config4 are higher than the L2 miss rates in config1 and config3.

Table. 2 gives the simulation results under Config3 without warming up the caches. These miss rates are directly collected by gem5. As we analyzed above, although the L2 miss rates of some benchmarks, e.g., 401, 403, 435 and 462, are very high, their absolute L2 accesses numbers are only tenth even hundredth of those of L1 accesses.

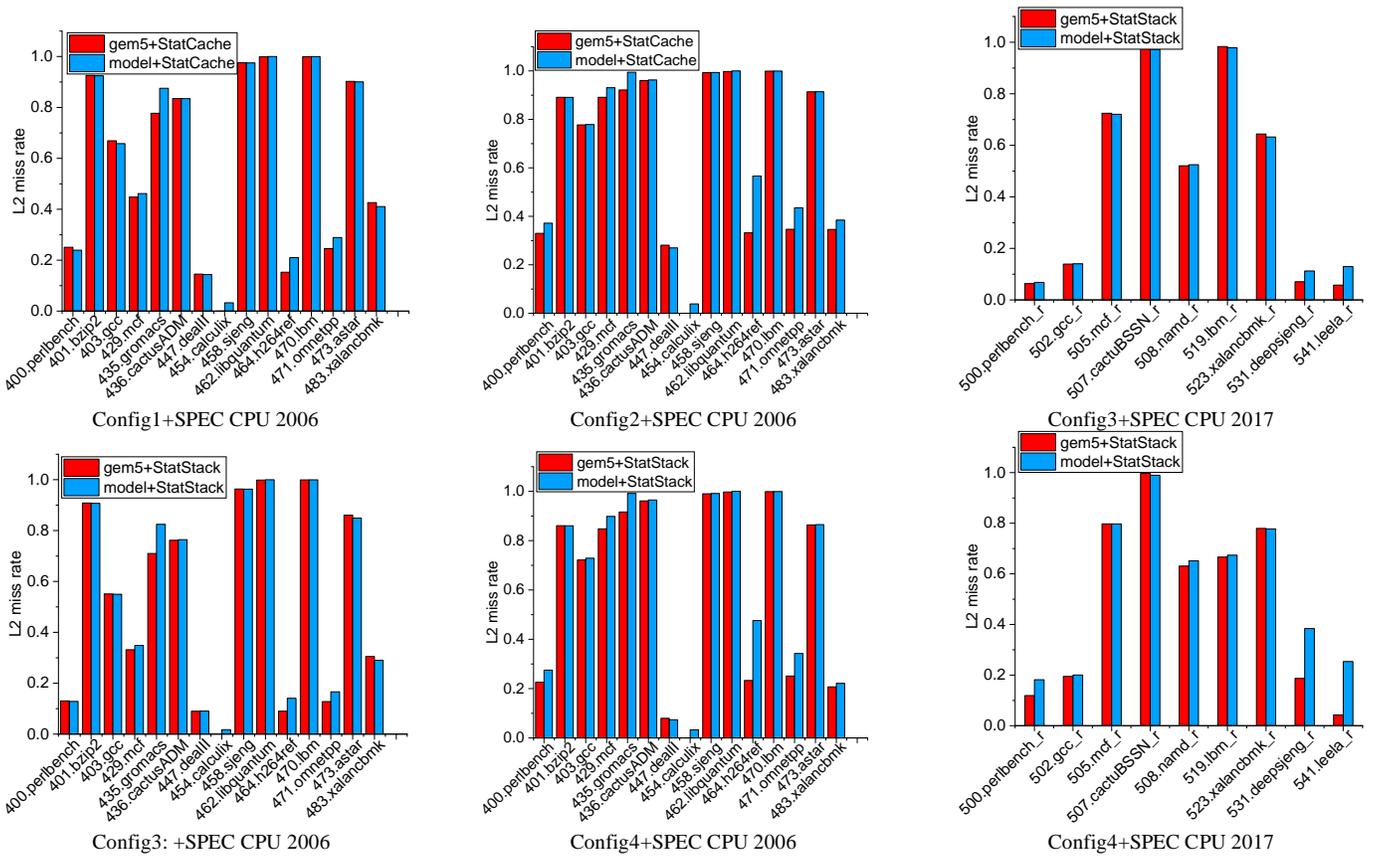

Fig. 14. Comparisons of L2 miss rates

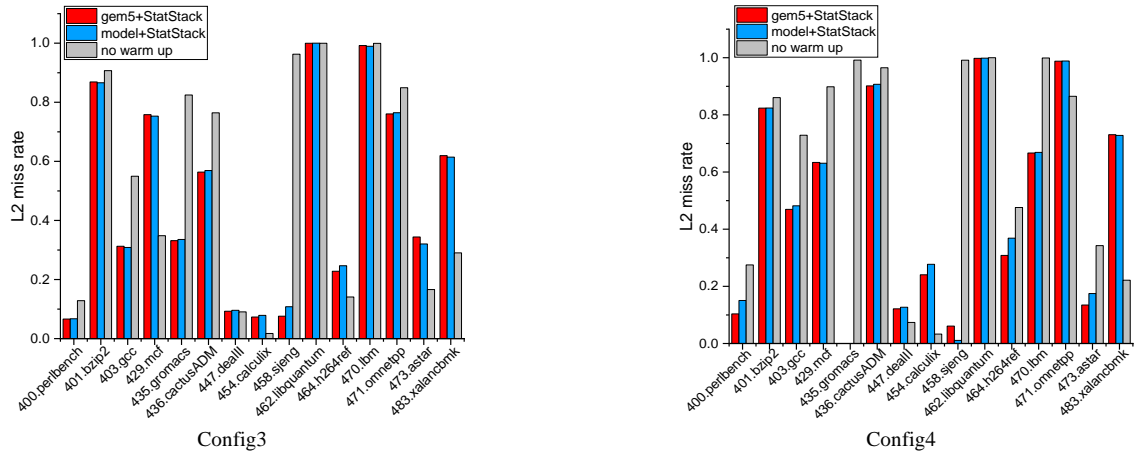

Fig. 15. Compassions of L2 miss rates with/without warming up

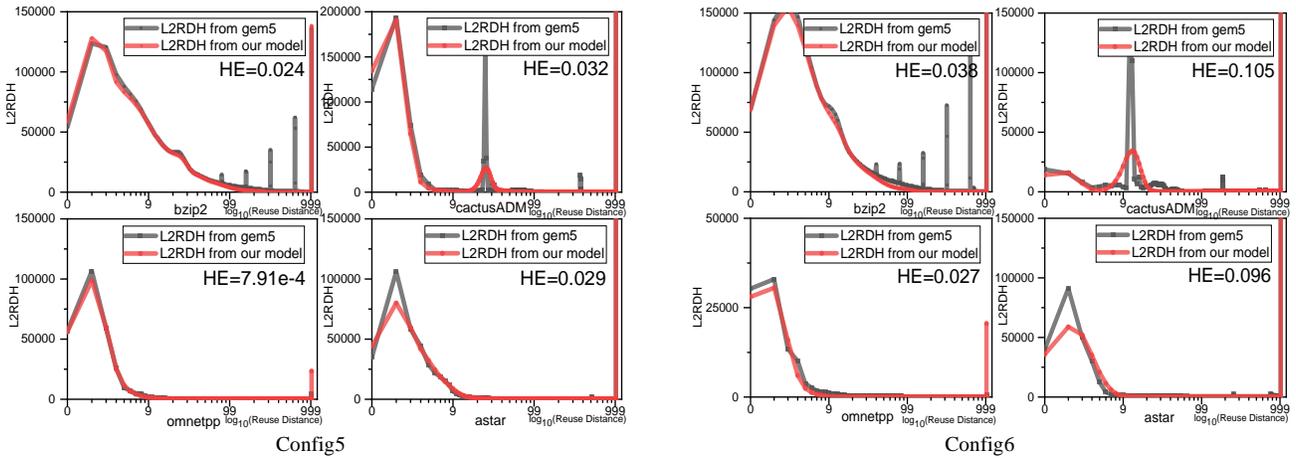

Fig. 16. Compassions of the L2 RDH between gem5 simulation results and our model

Table. 2 Simulation results directly from gem5 under Config3

|  | L1 accesses | L2 accesses | L1 misses | L2 misses | L1 miss rate | L2 miss rate |
|---|---|---|---|---|---|---|
| 400.perlbench | 4.97E+7 | 2.38E+6 | 2.38E+6 | 1.67E+5 | 0.05 | 0.07 |
| 401.bzip2 | 5.17E+7 | 4.22E+6 | 4.22E+6 | 3.89E+6 | 0.08 | 0.89 |
| 403.gcc | 4.71E+7 | 1.64E+6 | 1.64E+6 | 1.09E+6 | 0.03 | 0.66 |
| 429.mcf | 3.25E+7 | 2.94E+5 | 2.94E+5 | 1.07E+5 | 0.01 | 0.36 |
| 435.gromacs | 3.93E+7 | 3.66E+4 | 3.66E+4 | 2.89E+4 | 0.9E-3 | 0.79 |
| 436.cactusADM | 4.51E+7 | 1.42E+6 | 1.42E+6 | 1.13E+6 | 0.03 | 0.79 |
| 447.dealII | 2.37E+7 | 7.62E+6 | 7.62E+6 | 1.01E+6 | 0.32 | 0.13 |
| 454.calculix | 3.42E+7 | 1.76E+5 | 1.76E+5 | 8.17E+3 | 0.005 | 0.05 |
| 458.sjeng | 8.61E+7 | 8.68E+6 | 8.68E+6 | 8.37E+6 | 0.10 | 0.96 |
| 462.libquantum | 2.50E+7 | 7.81E+5 | 7.81E+5 | 7.81E+5 | 0.03 | 0.99 |
| 464.h264ref | 4.89E+7 | 6.93E+5 | 6.93E+5 | 1.34E+5 | 0.01 | 0.19 |
| 470.lbm | 7.43E+7 | 6.75E+6 | 6.75E+6 | 6.74E+6 | 0.09 | 0.99 |
| 471.omnetpp | 2.03E+7 | 3.22E+5 | 3.22E+5 | 1.18E+5 | 0.01 | 0.36 |
| 473.astar | 4.14E+7 | 1.68E+6 | 1.67E+6 | 1.41E+6 | 0.04 | 0.83 |
| 483.xalancbmk | 4.86E+7 | 2.92E+6 | 2.92E+6 | 1.60E+6 | 0.06 | 0.55 |

To verify our hypothesis that there are too many cold misses during the beginning of the system, we conduct experiments with 1 billion instructions to warm up the multi-level cache architecture. Then, we profile the trace information of the second 500 million instructions and evaluate the multi-level cache architecture again. The results are shown in Fig. 15. The grey bars mean the miss rates of the L2 cache without warming up the caches. We can see that after warming up the caches, the miss rates would decrease, which is caused by the decreasing of the cold misses. There are some exceptions in benchmarks of 429, 464, 471 and 483, which we believe are caused by the relatively small instruction counts in our study.

To evaluate our model under the cache architectures with different number of sets in the L1 cache and the L2 cache, we conduct experiments under four different cache configurations (Config5, Config6, Config7 and Config8). The ratios of sets of the L1 cache and the L2 cache are 1:2 or 1:4. Fig. 16 gives the comparisons of the L2 RDHs under the cache configurations with the set number ratio of 1:2 (Config5) and 1:4 (Config6). In Fig. 16, we can see that around the reuse distance of 30 in *cactusADM*, the results of our model smooth the characteristics of the L2 RDHs. In the tail of the L2 RDHs of *bzip2* in Fig. 16, there are some low peaks in the simulated curves, but the results from our model cannot reflect these subtle fluctuations. From the labeled *HEs* in each sub-figure of Fig. 16, we can also find the larger errors of the predicted RDH curves, which present an average accumulated error of each RDH bin of 4.33%. We believe that these errors are caused by the usage of the fixed probability of $P_{same}$. The actual $P_{same}$ of each reference may be different, which means $P_{same}$ should actually be a probability distribution instead of a fixed ratio (in our experiments, however, we consider $P_{same}$ as the fixed ratio of the number of sets in the L1 cache and the L2 cache). Fig. 17 gives the comparisons of L2 miss rates under the cache configurations with the set number ratio of 1:2 and 1:4. The average absolute errors of benchmarks from SPEC CPU 2006 under the four different cache configurations are 4.31%, 9.95%, 4.15% and 9.48%, respectively. The average errors of applications in SPEC CPU 2017 with the cache Config7 and Config8 are 4.70% and 5.79%.

The average error of all the results in the above experiments of calculating the miss rates is 4.76%. The main reason of the error is that we use a fixed ratio of $P_{same}$ in our model. Because the actual probability that two references in the same cache set in L1 cache are indexed to the same L2 cache sets is a probability distribution rather than a fixed ratio.

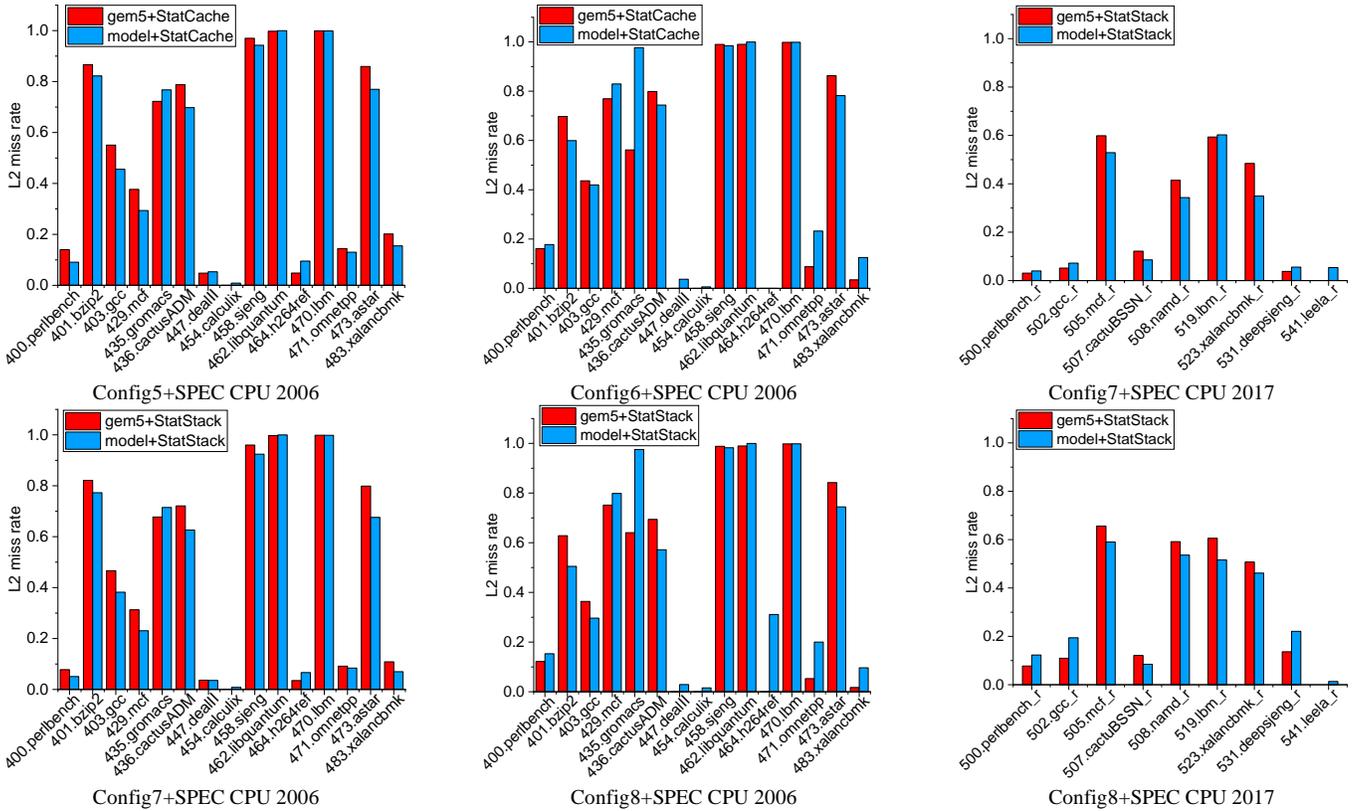

Fig. 17. Compassions of L2 miss rates

Table. 3 Hardware configurations of our profiling platform

| CPU | Intel(R) Core (TM) i7-4790 CPU @3.60GHz, 4 cores / 8 threads |
|---|---|
| RAM | 8GB DDR3 1600MHz ×2 |
| Storage | 1TB HDD SATA3 |
| OS | Ubuntu 18.04.1 LTS |

Table. 3 shows the hardware configurations of the profiling platform in our experiments. Fig. 18 shows the time overhead comparison of our model with gem5 detailed simulations. We compare four benchmarks under the same L1 cache configuration with four different L2 cache configurations (config1, config3, config5 and config7). In Fig. 18, the grey bars mean the evaluation time of our model for a single L2 cache configuration (meaning that the profiled L1 locality information only used for the evaluation of one L2 configuration and not re-used by other L2 candidates). Considering reusing of the profiling results, the average time consumption of each of the four different L2 caches can be evened to almost a quarter of the original time (red bars in Fig. 18). We can see that our model significantly reduces the evaluation time because the dominant time-consuming part, which is the profiling, can be completed much faster under AtomicSimpleCPU mode than detailed simulation in gem5. It is also worth to note that for each individual benchmark under each L2 cache configuration, traditional simulation-based method needs to re-run the simulation. In contract, for a given configuration of the L1 cache, the profiled results of our model can be reused in different L2 configurations. In K Ji's work [11], Eq. (12) is used to estimate the speeding up of the model. Because one-time profiling results can be reused for different L2 cache configurations, the profiling time should be divided by the number of L2 cache configurations. However, the premise of using Eq. (12) is that the time used for formulas calculation can be ignored. Unfortunately, we find the calculation complexity in K Ji's work is extremely high and the computing time cannot be ignored. The main computing equation of K Ji's work is shown in Eq. (13), which has an approximate complexity of $O(n^5)$ and possibly even higher when we considering the cases of the combination of $M$ and $N$. Thus, we argue the usage of Eq. (12) to evaluate the speed up of K Ji's work is not appropriate. On the contrary, the main equations of our model, i.e., Eq. (6) and Eq. (8), have the complexity of $O(n^2)$. Considering we set $n$ as 1024 (the cutting off reuse/stack distances are 1024 in our model), the computing workload of proposed model can be limited in an acceptable scale. In fact, it only takes several seconds to complete the calculation in our experiments when the required locality information has been profiled. Therefore, the speed up of our model can be accurately evaluated by Eq. (12), from which the average speed up of the evaluation for a single L2 configuration is about 7.38X and the average speed up of 4 L2 configurations can be almost 30X (7.38X*4). In comparison of the accuracies the average error of our model is 4.76%, which is almost same as the result of K Ji's work (5%) [11] and less than the result of another work (8%) [12].

$$Speeding\ up = \frac{gem5\ detailed\ simulation\ time}{\frac{L1\ profiling\ time}{L2\ cache\ configurations}} \quad (12)$$

$$L2RDH(K) = \sum H_{L1}(S) \times \sum\sum C_M^m (1-P_d)^m P_d^{M-m} \sum P'_x(n) \quad (13)$$

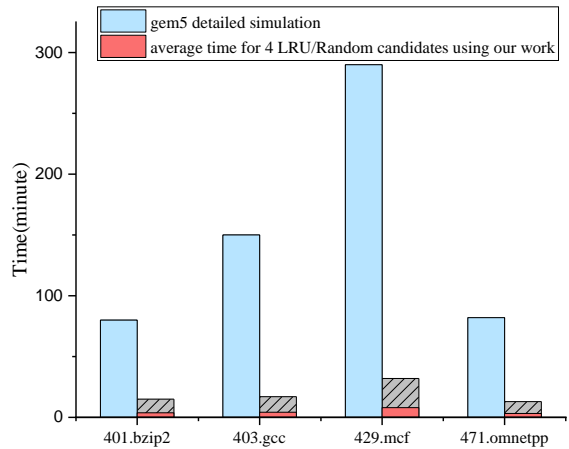

Fig. 18. Comparisons of Estimation Time

## VIII. CONCLUSION

In this paper, we propose two new metrics, RST table and Hit-RDH to describe more detailed information of the software traces. Helped by these two new metrics, we put forward a model, which takes the L1 RDH, RST table and Hit-RDH as the inputs and output the L2 RDH. Combined with StatCache and StatStack, our model can be applied to evaluate the L2 cache miss rate with Random or LRU replacement polices. The L1 cache RST table and Hit-RDH merely need to be profiled once for each benchmark for a given L1 cache architecture and they can be re-used for evaluations of different L2 cache configurations. Compared with the results from gem5 simulations, the average evaluation time for each L2 configuration can be sped up by almost 30X and the average absolute error is 4.76% for four different L2 cache candidates.

ACKNOWLEDGMENT

This work was supported by the Provincial Natural Science Foundation of Jiangsu Province under Grant No. BK20181141 and the National Natural Science Foundation of China under Grant 61974024.